\documentclass[twocolumn,pre,amsmath,amssymb]{revtex4}
\usepackage{graphicx}
\usepackage{dcolumn}
\usepackage{bm}

\begin{document}
\title{Mixed Reality States in a Bidirectionally Coupled Interreality System}
\author{Vadas Gintautas}
\email{vgintau2@uiuc.edu}
\author{Alfred W. H\"{u}bler}
\email{a-hubler@uiuc.edu}
\affiliation{Center for Complex Systems Research, Department of
Physics, University of Illinois at Urbana-Champaign, Urbana,
Illinois 61801}
\date{\today}
\begin{abstract}
We present experimental data on the limiting behavior of an interreality system comprising a virtual horizontally driven pendulum coupled to its real-world counterpart, where the interaction time scale is much shorter than the time scale of the dynamical system.  We present experimental evidence that if the physical parameters of the simplified virtual system match those of the real system within a certain tolerance, there is a transition from an uncorrelated dual reality state to a mixed reality state of the system in which the motion of the two pendula is highly correlated.  The region in parameter space for stable solutions has an Arnold tongue structure for both the experimental data and for a numerical simulation.  As virtual systems better approximate real ones, even weak coupling in other interreality systems may produce sudden changes to mixed reality states.
\end{abstract}
\maketitle
Although increasingly sophisticated real-time computer simulations of the real world are created every day, to date there has been little to no research done on the physics of the pairing of a virtual system and its real-world counterpart, often referred to as an ``interreality'' system~\cite{kokswijk03}.  Virtual systems are often created to model real systems as accurately as possible, with great pains taken to eke out additional realism.  Aside from virtual worlds designed for entertainment, examples of accurate virtual models of the real world abound in high-energy physics accelerator work~\cite{quiang00b}.  Furthermore, a computer simulation can feature unidirectional coupling to the real world, as in the case of black box trading in finance~\cite{bennell04} or the dynamic clamp in neuroscience~\cite{prinz04}.  The unidirectional coupling can also be from the real world to the simulation, as in the case of data-driven modeling whereby live measurement data is incorporated into an executing application.  This has been used for creating accurate, real-time models of systems ranging from complex vortex flows~\cite{rossberg04} to human cancer cells~\cite{christopher04}.  The next step is to examine an interreality system in which there is bidirectional coupling.  
\begin{figure}[ht]
  \centering
     \includegraphics[width=0.35\textwidth]{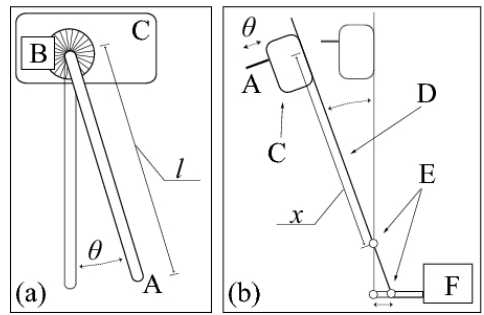} 
	\caption{Experimental apparatus. Fig.~\ref{fig:virtpend}(a) shows a side view detail of the pendulum housing: pendulum (A) is attached to slotted disk of angular encoder (B) in housing (C).  Fig.~\ref{fig:virtpend}(b) shows a top view of the apparatus: pendulum (A) in housing (C) is attached to lever arm (D) with pivot points (E), driven by actuator (F).  The pendulum has a length of $l=15.37$ cm and the adjustable distance $5.1$ cm $\le x\le 25$ cm from the pendulum to the pivot point controls the strength of the dimensionless coupling constant $0.17\le a_r\le 0.67$.  As indicated, $\theta$ is the angular displacement of the real pendulum as measured from the vertical.}
		\label{fig:virtpend}
\end{figure}

In this Brief Report we present experimental evidence for a transition from a dual reality state to a mixed reality state in an interreality system featuring bidirectional, instantaneous coupling.  The experimental phase diagram is in the form of an Arnold tongue~\cite{jensen02}.  There is good agreement between the experimental data and simulation.  We demonstrate that even a simple model taking into account a single degree of freedom and using only linear damping is sufficient to give rise to this mixed reality state.  It is often difficult and prohibitively expensive to create a model that predicts all observable parameters of a system in the real world to maximum precision.  Even a computer model of the familiar physical pendulum in air must take into account linear and quadratic damping, the buoyancy of the pendulum in the air, the added mass due to the pendulum dragging air with it as it moves, and a half dozen other effects in order to reproduce the measured period to a precision of $10^{-5}$ seconds~\cite{nelson86}.  All but the most sophisticated computer simulations will use approximations when it comes to modeling a physical system. 

As an example of a virtual system coupled to a real one, we choose a horizontally driven physical pendulum as the real system.  The horizontally and vertically driven physical pendulum has been described in the literature~\cite{phelps65}.  The known equations of motion accurately model the dynamics of the system; to create a physically accurate virtual pendulum we need only to make a real-time simulation based on these equations.  The experimental physical pendulum is a lightweight, very thin wooden rod with a length $l = 15.37$ cm, a diameter of $2.3$ mm, and a mass of $0.4$ g.  The pendulum is connected directly to the roller of a digital angular encoder.  The data from the angular encoder is sent to a computer; this provides a simple and robust measure of the instantaneous angular position of the pendulum.  Forcing for the physical pendulum is provided by an amplifier and a PASCO actuator (model SF-9324) with a maximum displacement of $0.3$ cm.  The pendulum is attached to a lever arm that allows it to be driven with a greater amplitude $x_{drive}$, up to a maximum of $5.75$ cm.  The computer controls the voltage to the actuator via an analog DAQ.  Figs.~\ref{fig:virtpend}(a) and~\ref{fig:virtpend}(b) show the experimental setup.

\begin{figure}[ht]
  \centering
     \includegraphics[width=0.35\textwidth]{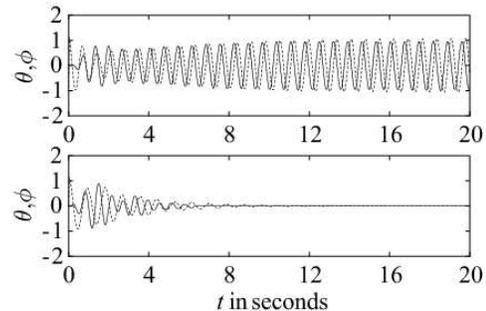} 
	\caption{The top plot shows position versus time for the experimental data point $\bar{\omega}=0.98$, $\bar{a}\equiv\sqrt{a_v a_r}=0.19$ (Region I, mixed reality state).  This plot is an example of the system exhibiting stable, phase-locked periodic motion.  The bottom plot has $\bar{\omega}=0.72$ and $\bar{a}=0.22$ (Region II, dual reality state).  This is an example of the system ending in the stable equilibrium position $\phi=\theta=0$. For both plots, the solid and dashed lines correspond to $\theta$ (the position of the real pendulum) and $\phi$ (the position of the virtual pendulum), respectively.}
	\label{fig:twoplot}
\end{figure}

We calculate the equation of motion for a horizontally driven physical pendulum in the usual way.  See Phelps and Hunter for a detailed derivation~\cite{phelps65}.  We define $\theta$ as the angle through which the real pendulum oscillates as measured from the vertical, and we define $\phi$ to be the corresponding angle for the virtual pendulum.  Then the equation of motion for the horizontally driven virtual pendulum with an arbitrary time-dependent driving function $f$ is 
\begin{equation}
\label{eq:virtpendeqn}
\ddot{\phi}+2\beta\dot{\phi}+\left(\bar{\omega}\omega_0\right)^2\sin{\phi}+a_v\ddot{f}_v\cos{\phi}=0,
\end{equation}
where $\beta$ is the damping coefficient, $\omega_0\equiv2\pi\omega_r$ is the natural angular frequency of the real pendulum, $\bar{\omega}\equiv\omega_v/\omega_r$ is the dimensionless ratio of the natural frequencies of the two pendula.  In terms of the pendulum moment of inertia $I$ and mass $m$, we define $a\equiv I^{-1}mlx_{drive}\propto$~$x_{drive}/l$ to be the dimensionless coupling constant that sets the scale of the coupling term.  Weak coupling corresponds to $a \ll 1$.  To determine the coefficient of linear damping, we fit the decay of the uncoupled real pendulum to a simple exponential.  For our real pendulum this measured value of $\beta$ is $0.45 \pm 0.2$, while the measured natural frequency of the real pendulum $\omega_{r}$ is $1.57\pm0.01 s^{-1}$.  These are the values used for $\beta$ and $\omega_{r}$ in each calculation.  In this work, an overbar on a variable denotes a normalized, dimensionless quantity.  Also, the subscripts $v$ and $r$ refer to variables associated with the virtual pendulum and the real pendulum, respectively. A standard fifth-order Runge-Kutta routine is used to integrate this equation.  We let $f$ scale with the angular displacement of the real pendulum: $f_v(\theta)\equiv \theta$.  At the $n$-th time step of duration $\Delta t$, the measured positions of the real pendulum at the current and two previous time steps determine the value of $\ddot{f}$:
\begin{equation}
\label{eq:evalfdotdot}
\ddot{f}_{v,n}=\left(\frac{\theta_{n}-2\theta_{n-1}+\theta_{n-2}}{\Delta t^2}\right)
\end{equation}
Using these values, the integrator returns $\phi_{n}$, which is used to determine the driving amplitude $f_{r,n}\equiv \phi_{n}$ for the real pendulum.  A voltage proportional to $f_{r,n}$ is sent to the actuator driving the real pendulum.  The distance from the pivot point along the lever arm determines $a_r$ [see Fig.~\ref{fig:virtpend}(b)].  Since the measurement, integration, and feedback are easily completed by the computer in a time $\delta t_{computer}$ with $\delta t_{computer} < \Delta t \ll 1/\omega_0$, the program then waits to integrate again until the internal timer reaches $n\Delta t$ after the initial starting time for the first integration.  Since it is impractical to attempt to release the real pendulum from precisely the same starting position each time the program is run, instead the virtual pendulum is started with zero initial velocity but with a nonzero initial position.  The real pendulum is started at rest, with $\theta_{n=0}=0$.  Two typical sets of position versus time data are plotted in Fig.~\ref{fig:twoplot}.

With $\Delta t$ = 35 ms, the bidirectional feedback is effectively instantaneous; the effect is that of a real-time virtual pendulum simulation that immediately responds to any motion of the real pendulum and vice versa.  This works because $\Delta t$ is much shorter than the characteristic time scales of the dynamical system.  The natural frequency of the real pendulum is approximately $1.57 s^{-1}$.  With no feedback, the motion of either pendulum ceases after less than $10$ s due to friction in the real pendulum and damping in the equation of motion of the virtual pendulum.  We allow the system to run for $45$ s, long enough for any transient dynamics to vanish, and then we evaluate the final dynamics of the system.  We find that there are two equilibrium states of the system when $a_v$ and $a_r$ are restricted to a range appropriate to weak coupling $(a_v, a_r <0.4)$.  We label these the ``dual reality'' state and the ``mixed reality'' state.  In the dual reality state, the oscillations of both pendula are uncorrelated and decrease in amplitude until both come to rest at the stable position $\phi=\theta=0$.  The two pendula behave as separate oscillators in the dual reality state, with reality and virtual reality interacting but coexisting individually. In the mixed reality state, the two pendula exhibit highly correlated stable, phase-locked periodic motion.  In this mixed reality state of the system, the real pendulum and the virtual pendulum move together as one.  These are illustrated in Fig.~\ref{fig:twoplot}.

We model this coupled system by removing the real pendulum entirely and replacing it with a routine in the code that separately integrates 
\begin{equation}
\label{eq:realpendeqnsim}
\ddot{\theta}_{sim}+2\beta\dot{\theta}_{sim}+\omega_{0}^2\sin{\theta_{sim}}-a_r\ddot{f_r}\cos{\theta_{sim}}=0,
\end{equation}
an independent and equivalent equation of motion that represents the real pendulum.  The $(-)$ sign with the $a_r$ term is necessary because the lever arm is mounted on a pivot that effectively reverses the direction of the movement of the actuator [see Fig.~\ref{fig:virtpend}(b)].
Since only the position of the real pendulum is measured in the experimental setup, the integration of Eq.~(\ref{eq:realpendeqnsim}) returns the position $\theta_{sim}$.  The velocity is calculated using the position at the previous time step, and $\ddot{f}_r$ is evaluated analogously to Eq.~(\ref{eq:evalfdotdot}).  The virtual pendulum routine independently integrates 
\begin{equation}
\label{eq:virtpendeqnsim}
\ddot{\phi}_{sim}+2\beta\dot{\phi}_{sim}+\left(\bar{\omega}\omega_0\right)^2\sin{\phi_{sim}}+a_v\ddot{f}_v\cos{\phi_{sim}}=0
\end{equation}
at each time step.  The only difference between the experimental system and the simulation system is that instead of measuring the position of the real pendulum, the simulation integrates an equation of motion to calculate that position.  The feedback works exactly the same way as before, except instead of an output voltage the program simply provides a feedback amplitude coefficient to the modeled real pendulum's integrator at each time step.  This simulation can also run at real time, but suppressing the delay between integrations returns precisely the same results in a fraction of the time required.

\begin{figure}[ht]
  \centering
     \includegraphics[width=0.35\textwidth]{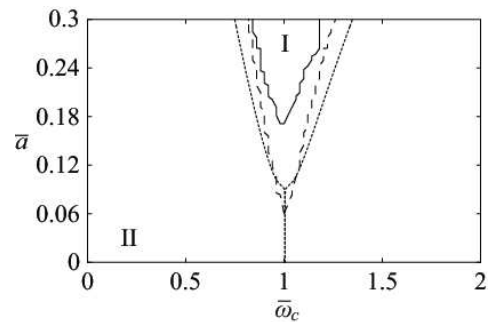} 
	\caption{Limiting behavior phase transition diagram with Arnold tongue structure for the two parameters $\bar{a}$ and $\bar{\omega}$.  The solid, dashed, and dotted lines indicate the critical points $\bar{\omega}_{c}$ in the experiment, simulation, and linear theory, respectively.  These critical points are the boundary between Regions I and II in this parameter space.  Region I corresponds to stable, phase-locked oscillations of the center of mass of each pendulum; this is the mixed reality state.  Region II corresponds to both pendula ending at the stable equilibrium position $\phi=\theta=0$; this is the dual reality state.}
	\label{fig:paraspace}
\end{figure}

\begin{figure}[ht]
  \centering
     \includegraphics[width=0.35\textwidth]{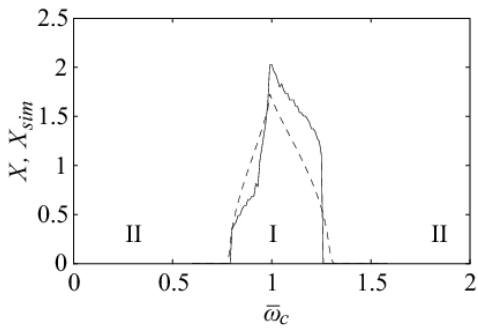} 
	\caption{This figure shows the response amplitudes $X$ and $X_{sim}$ versus $\bar{\omega}$.  The solid and dashed lines correspond to the experimental and simulation data, respectively.  These are response curves for the system with $\bar{a}=0.364$.  As indicated, for the experiment, the interval in $\bar{\omega}$ where $X>0$ corresponds to Region I, while the interval in $\bar{\omega}$ where $X=0$ corresponds to Region II.  Likewise, for the simulation, the interval in $\bar{\omega}$ where $X_{sim}>0$ corresponds to Region I, while the interval in $\bar{\omega}$ where $X_{sim}=0$ corresponds to Region II (see Fig.~\ref{fig:paraspace}).}
	\label{fig:rescurves}
\end{figure}

We now work with the reduced parameter space described by $\bar{\omega}$ and $\bar{a}\equiv\sqrt{a_v a_r}$, where $\bar{a}$ is the geometric mean of the forcing amplitudes.  $\bar{a}$ characterizes the strength of the bidirectional coupling; necessarily $\bar{a}\rightarrow0$ as $a_v\rightarrow0$ or $a_r\rightarrow0$.  For weak coupling, we have $0\leq\bar{a}<0.4$.  $\bar{\omega}$ characterizes the quality of the model.  For the virtual pendulum to be an accurate model of the real pendulum, $\bar{\omega}$ has to be near 1.  We find that there are two distinct regions in this parameter space, corresponding to two qualitatively different limiting behaviors of the system.  These are depicted in Fig.~\ref{fig:paraspace}. Region I corresponds to the mixed reality state of the system.  The oscillations are about the fixed point of each pendulum, and occur at frequencies close to the natural frequency of the real pendulum (see the top plot in Fig.~\ref{fig:twoplot}).  Region II corresponds to the dual reality state of the system.  In this region, both pendula initially oscillate but the system is unable to sustain this uncorrelated motion and loses kinetic energy until both pendula are at rest (see the bottom plot in Fig.~\ref{fig:twoplot}).  Region I has the Arnold tongue structure for mode-locked solutions in parameter space, as seen in Fig.~\ref{fig:paraspace}.  For each data set we wait until $t=25$ s, which is long enough for the transient dynamics to vanish, then we measure the maximum amplitude of the real pendulum $X\equiv\max{(\theta)}$ over several oscillation cycles for each pair of parameters $\bar{\omega}$ and $\bar{a}$.  We define $X$ as the maximum amplitude of the real pendulum in the experimental system and $X_{sim}$ as the maximum displacement of the simulated real pendulum.  $X$ and $X_{sim}$ are plotted against $\bar{\omega}$ for $\bar{a}=0.364$ in Fig.~\ref{fig:rescurves}.  This curve shows the phase transitions from Region II to Region I, then back to Region II.

Dropping the \textit{sim} subscripts, to linear order Eqs.~(\ref{eq:realpendeqnsim}) and~(\ref{eq:virtpendeqnsim}) become
\begin{eqnarray}
\ddot{\theta}+2\beta\dot{\theta}+\omega_{0}^{2}\theta-a_{r}\ddot{\phi}&=&0,\label{eq:linearsystemreal}\\
\ddot{\phi}+2\beta\dot{\phi}+\bar{\omega}^{2}\omega_{0}^{2}\phi+a_{v}\ddot{\theta}&=&0.\label{eq:linearsystemvirt}
\end{eqnarray}
By taking successive derivatives of Eq.~(\ref{eq:linearsystemreal}) and substituting these into the second derivative of Eq.~(\ref{eq:linearsystemvirt}), this system can be written as the decoupled linear fourth-order differential equation $(1+\bar{a}^{2})\theta^{(4)}+4\beta\theta^{(3)}+(4\beta^{2}+\gamma^{2})\ddot{\theta}+2\beta\gamma^{2}\dot{\theta}+\bar{\omega}^{2}\omega_{0}^{4}\theta=0$.  $\theta^{(4)}=\theta^{(3)}=\ddot{\theta}=0$ at $t=0$ comprise the initial conditions, and we define $\gamma\equiv\omega_{0}^{2}(1+\bar{\omega}^{2})$.  The general solution to this system is in the form $\theta(t)=\sum_{i=1}^{4}c_{i}e^{\lambda_{i}t}$, where the $c_{i}$ are constants determined from the initial conditions and the $\lambda_{i}$ are the four solutions of the characteristic equation $(1+\bar{a}^{2})r^{4}+4\beta r^{3}+(4\beta^{2}+\gamma^{2})r^{2}+2\beta\gamma^{2}r+\bar{\omega}^{2}\omega_{0}^{4}=0$.  If, for all $i$, $\lambda_{i}<0$, then we have $\theta(t)\rightarrow 0$ as $t\rightarrow\infty$.  However, even if for one eigenvalue we have $\text{Re}(\lambda_{i})>0$, then the solution is no longer bounded: $\theta(t)\rightarrow 0$ as $t\rightarrow\infty$.  For a given $\bar{a}$, there are two values for $\bar{\omega}$ such that $\text{Re}(\lambda_{i})=0$ for one eigenvalue.  We define these to be the critical points $\bar{\omega}_{c}$ that mark the boundaries of the phase diagram in Fig.~\ref{fig:paraspace}.  As shown in this figure, the phase diagram of the linear system in Eqs.~(\ref{eq:linearsystemreal}) and~(\ref{eq:linearsystemvirt}) closely matches that of the full system in Eqs.~(\ref{eq:realpendeqnsim}) and~(\ref{eq:virtpendeqnsim}).

There are differences between the experimental and simulation data.  Region I in the simulation data extends further in the direction of small $\bar{a}$ than in the experimental data (Fig.~\ref{fig:paraspace}).  Since the general shape of Region I is very similar for both sets of data, this onset of phase-locked oscillations for smaller $\bar{a}$ appears to reflect the greater efficiency of the simulation.  In the real system, there are inevitable small delays and noise in the electronics and computer control, as well as additional friction terms beyond the linear damping term in Eq.~(\ref{eq:realpendeqnsim}).  As discussed above, more sophisticated models of friction are necessary to reproduce pendulum dynamics accurately.  Nonetheless, the virtual pendulum in Eq.~(\ref{eq:virtpendeqn}) sufficiently models the real pendulum, giving rise to the mixed reality state when $\bar{\omega}\approx 1$.

This work presents experimental data from an interreality system comprising a virtual pendulum and its real-world counterpart.  There is bidirectional, instantaneous coupling between the two pendula.  We find that if the dynamics of the virtual system approximate those of the real one within a small tolerance, there is a phase transition in the behavior of the system.  The interreality system makes a transition from a dual reality state in which reality and virtual reality are uncorrelated to a mixed reality state in which reality and virtual reality are highly correlated.  For both the experimental data and a numerical simulation, the region of mixed reality mode-locked solutions in parameter space is an Arnold tongue.  While the appearance of an Arnold tongue is not surprising, it highlights two features of this interreality system that we would expect to be present in similar systems.  The shape of the Arnold tongue indicates that with stronger coupling, the mixed reality states are accessible even with increased mismatch between the frequencies of the pendula.  Also, the boundary of an Arnold tongue represent a discontinuous or sudden change change from a dual reality state to a mixed reality state, as seen in Fig.~\ref{fig:rescurves}.  As virtual systems better approximate real ones, even weak coupling in other interreality systems may produce sudden changes to mixed reality states.

Forced systems tend to have the greatest response when the dynamics of the driving function match the dynamics of the system~\cite{lai05}.  This has been studied in depth for various damped, coupled oscillator systems~\cite{arsenault95}.  One application of this phenomenon is resonance spectroscopy, where a system with an unknown parameter $b$ is driven with a forcing function that depends on this parameter.  As $b$ is varied, the system has the largest response when the dynamics of the forcing function match those of the system; thereby the value of $b$ is identified~\cite{wargitsch95b}. To date, work in this area has predominantly focused on numerical analysis of coupled differential equations that represent the equations of motion for linear and nonlinear oscillators~\cite{arsenault95,wargitsch95b}.  There has been some experimental work on non-sinusoidal driving of nonlinear oscillators~\cite{lai05}, but to the authors' knowledge there has been no work done on resonance spectroscopy via the instantaneous coupling of a virtual system to its real counterpart.  It may be possible in this system or similar systems to use either the peak of the response curve or the boundary of the Arnold tongue to do system identification.  Furthermore, it may be possible to experimentally observe universal routes to synchronization~\cite{taherion99, rosenblum97} using interreality systems.  We plan to examine these topics in the future.  This work was supported by the National Science Foundation Grant No. NSF PHY 01-40179, NSF DMS 03-25939 ITR, and NSF DGE 03-38215.


\begin{thebibliography}{10}

\bibitem{kokswijk03}
J. van Kokswijk, {\em Human, Telecoms \& Internet as Interface to Interreality}
  (Bergboek, The Netherlands, 2003).

\bibitem{quiang00b}
J. Qiang, R.~D. Ryne, S. Habib, and V. Decyk, J.\ Comput.\ Phys. {\bf 163},
  434  (2000).

\bibitem{bennell04}
J. Bennell and C. Sutcliffe, Int.\ J.\ of Intel.\ Sys.\ in Acct.,\ Fin.\ and
  Mgmt {\bf 12},  243  (2004).

\bibitem{prinz04}
A.~A. Prinz, L.~F. Abbott, and E. Marder, TRENDS in Neuroscience {\bf 27},  218
   (2004).

\bibitem{rossberg04}
A.~G. Rossberg, K. Bartholom\'{e}, and J. Timmer, Phys.\ Rev.\ E {\bf 69},
  016216  (2004).

\bibitem{christopher04}
R. Christopher, et al., Ann.\ N.\ Y.\ Acad.\ Sci. {\bf 1020},  132
  (2004).

\bibitem{jensen02}
R.~V. Jensen, Am.\ J.\ Phys. {\bf 70},  607  (2002).

\bibitem{nelson86}
R.~A. Nelson and M.~G. Olsson, Am.\ J.\ Phys. {\bf 54},  112  (1986).

\bibitem{phelps65}
I. F.~M.~Phelps and J. J.~H.~Hunter, Am.\ J.\ Phys. {\bf 33},  285  (1965).

\bibitem{lai05}
Y.-C. Lai, A. Kandangath, S. Krishnamoorthy, J.~A. Gaudet, and A.~P.~S.
  de~Moura, Phys.\ Rev.\ Lett. {\bf 94},  214101  (2005).

\bibitem{arsenault95}
L.~E. Arsenault and A.~W. H{\"u}bler, Phys.\ Rev.\ E {\bf 51},  3561  (1995).

\bibitem{wargitsch95b}
C. Wargitsch and A. H{\"u}bler, Phys.\ Rev.\ E {\bf 51},  1508  (1995).

\bibitem{taherion99}
S. Taherion and Y.-C. Lai, Phys.\ Rev.\ E {\bf 59},  R6247  (1999).

\bibitem{rosenblum97}
M.~G. Rosenblum, A.~S. Pikovsky, and J. Kurths, Phys.\ Rev.\ Lett. {\bf 78},
  4193  (1997).

\end{thebibliography}
\end{document}